\documentclass[9pt,twocolumn]{extarticle}
\pdfoutput=1

\usepackage{soul}
\usepackage{titlesec} 
\usepackage[backend=bibtex,style=nature]{biblatex}
\DeclareNameAlias{sortname}{first-last}
\DeclareNameAlias{default}{first-last}
\bibliography{biblio} 

\usepackage{dsfont}
\usepackage{graphicx}
\usepackage{subcaption}
\usepackage[margin=1in]{geometry}
\usepackage[usenames,dvipsnames]{color}
\usepackage[colorlinks,linkcolor=Blue,urlcolor=Blue,citecolor=Blue]{hyperref}
\usepackage{amsmath,amssymb}
\usepackage[squaren]{SIunits}

\usepackage{mathpazo} 
\usepackage{courier} 
\normalfont
\usepackage[T1]{fontenc}
\usepackage{caption}
\usepackage{upgreek}

\newcommand{\ad}[1]{\textsuperscript{#1}\kern-2pt}

\def\({\left(}
\def\){\right)}
\def\[{\left[}
\def\]{\right]}

\usepackage{capt-of}

\usepackage{flushend}

\def\({\left(}
\def\){\right)}
\def\[{\left[}
\def\]{\right]}

\def\mytitle{Experimental Quantum Hamiltonian Learning}   

\title{\vspace{-1.5cm}\Huge\textbf{\textrm{\mytitle}}} 

\author{Jianwei Wang$^{1\dagger\star}$, Stefano Paesani$^{1\dagger}$, Raffaele Santagati$^{1\dagger}$, Sebastian Knauer$^{1}$, Antonio A. Gentile$^{1}$, Nathan Wiebe$^{2\star}$, \\Maurangelo Petruzzella$^{3}$, 
Jeremy L. O\textquoteright Brien$^{1}$, 
John G. Rarity$^{1}$, 
Anthony Laing$^{1}$,
and Mark G. Thompson$^{1\star}$}

\date{} 
\begin{document}
\twocolumn[{%
\maketitle 
\vspace{-8mm}
\begin{center}
\begin{minipage}{0.95\textwidth}
\begin{center}
\textit{\textrm{
\textsuperscript{1} 
Quantum Engineering Technology Labs, H. H. Wills Physics Laboratory and Department of Electrical and Electronic Engineering, University of Bristol, BS8 1FD, UK. 
\\\textsuperscript{2} Quantum Architectures and Computation Group, Microsoft Research, Redmond, Washington 98052, USA.  
\\\textsuperscript{3} Department of Applied Physics, Eindhoven University of Technology, P.O. Box 513, NL-5600MB Eindhoven,
The Netherlands
\\\textsuperscript{$\dagger$} These authors contributed equally to this work 
\\\textsuperscript{$\star$} e-mails: jianwei.wang@bristol.ac.uk; nawiebe@microsoft.com; mark.thompson@bristol.ac.uk 
\\ 
}}
\end{center}
\end{minipage}
\end{center}
}]

\noindent
\textbf{
Efficiently characterising quantum systems~\cite{Cram10efficient, SilvaPRL,spagnolo2016},
verifying operations of quantum devices~\cite{Spagnolo:2014, Carolan:NP, Barz:2013}
and validating underpinning physical models~\cite{Parsons1253, Labuhn:2016}, 
are central challenges for the development of quantum technologies~\cite{Carolan:15,Debnath2016, Barends}
and for our continued understanding of foundational physics~\cite{Sh-NatPhys-10-278, Arndt:2014bp}. 
Machine-learning enhanced by quantum simulators has been proposed as a route to improve the computational cost of performing these studies~\cite{Wiebe2014, Wiebe2014b}. 
Here we interface two different quantum systems through a classical channel --- a silicon-photonics quantum simulator and an electron spin in a diamond nitrogen-vacancy centre --- and use the former to learn the latter's Hamiltonian via Bayesian inference. 
We learn the salient Hamiltonian parameter with an uncertainty 
of approximately $10^{-5}$. 
Furthermore, an observed saturation in the learning algorithm suggests deficiencies in the underlying Hamiltonian model, which we exploit to further improve the model itself. 
We go on to implement an interactive version of the protocol and experimentally show its ability to characterise the operation of the quantum photonic device. 
This work demonstrates powerful new quantum-enhanced techniques for investigating foundational physical models and characterising quantum technologies. 
}

In science and engineering \cite{williams1996transmission, zewail2000femtochemistry},
physical systems are approximated by simplified models to allow the comprehension of their essential features. The utility of the model hinges upon the fidelity of the approximation to the actual physical system, and can be measured by the consistency of the model predictions with the real experimental data. 
However, predicting behaviour in the exponentially large configuration space of quantum systems is known to be intractable to classical computing machines~\cite{Feynman1982, Lloyd1996}. A fundamental question therefore naturally arises: How can underpinning theoretical models of quantum systems be validated?   

\begin{figure*}[ht!]
\centering
\includegraphics[width=1\textwidth]{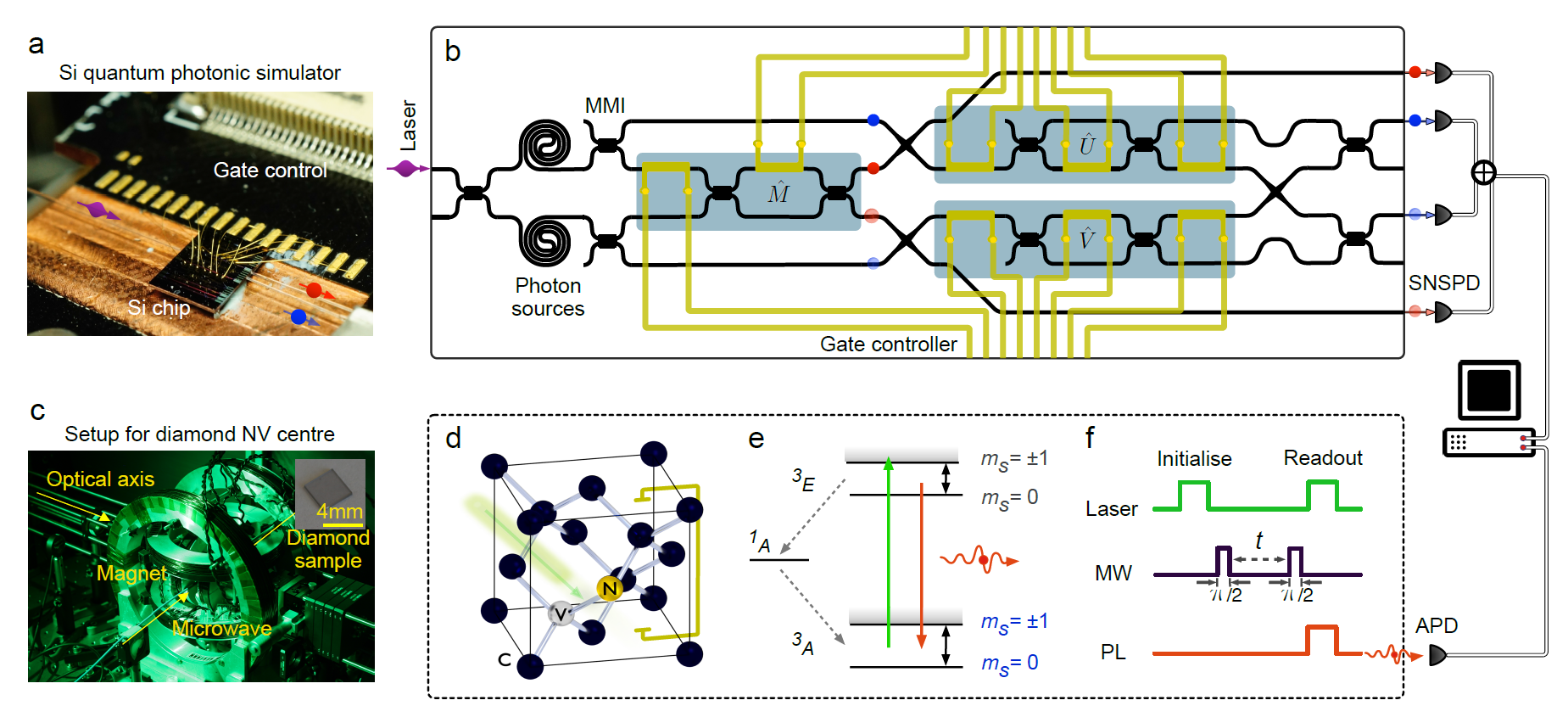}
\caption{\textbf{Quantum photonic simulator and diamond nitrogen-vacancy centre}.  
The silicon quantum photonic simulator: \textbf{a}, experimental setup and \textbf{b}, device schematic. 
A $10$ mW continuous-wave pump laser with near $1550$ {\nano\meter} wavelength is coupled into the chip through optical fibres.  
In (\textbf{b}), black lines are silicon nano-photonic waveguides, and gold wires represent thermo-optical phase-shifters and their transmission lines.    
A pair of idler (blue) and signal (red) photons with different wavelengths are generated via spontaneous four-wave mixing in the spiral waveguide sources. 
These photons are split equally via multi-mode interferometer (MMI) beam-splitters, producing a post-selected maximally entangled state. 
The operation $\hat{U}$ or $\hat{V}$ performed on idler qubit is coherently controlled by the state of the signal qubit.  
Performing measurements $\hat{M}$ on the signal qubit allows to estimate the likelihood function for the chosen configuration of the device. 
Photons are detected off-chip by fibre-coupled superconducting nanowire single-photon detectors (SNSPD). 
\textbf{c}, Confocal setup with diamond (inset) containing NV$^-$ centres.
\textbf{d}, Structure and \textbf{e}, energy level diagram of an NV$^{-}$ centre in diamond. 
The ground-state electron spin Hamiltonian, describing the coherent dynamics between $m_\text{s}=0$ and $-1$, is to be characterised and learned using the quantum simulator in (\textbf{a}). 
\textbf{f}, A single Rabi sequence for the initialisation, manipulation and read-out of the electron spin state. 
A laser pulse at $532$ {\nano\meter}  is used to initialise the spin into $m_\text{s}=0$. Two microwave (MW) $\pi/2$-pulses with a time $t$ delay are then used to coherently drive the spin. The spin state is measured by detecting photo-luminescence (PL) with an avalanche photodiode (APD). 
These two different physical systems are interfaced through a classical computer. }
\label{fig:NVvsChip}
\end{figure*}

To address this question, \textit{quantum Hamiltonian learning} (QHL) was recently proposed ~\cite{Wiebe2014,Wiebe2014b} as a technique that exploits classical machine learning  with quantum simulations to efficiently validate Hamiltonian models and verify the predictions of quantum systems or devices.
QHL is tractable in cases in which other known methods fail
because quantum simulation is exponentially faster than existing techniques~\cite{Feynman1982, Lloyd1996,AspuruGuzik2005} for simulating broad classes of complex quantum systems~\cite{Kim2010,PeruzzoChem, Lanyon2010,OMalley2016}.
Our experimental demonstration of QHL uses a programmable silicon-photonics quantum simulator, shown in Figs.~\ref{fig:NVvsChip}a,b, to learn the electron spin dynamics of a negatively charged nitrogen-vacancy (NV$^-$) centre in bulk diamond, shown in Figs.~\ref{fig:NVvsChip}c,d.
We further demonstrate an interactive QHL protocol that allows us to characterise and verify single-qubit gates using other trusted gates on the same quantum photonic device. 
  
\begin{figure*}[ht!]
	\centering
 \includegraphics[width=1\textwidth]{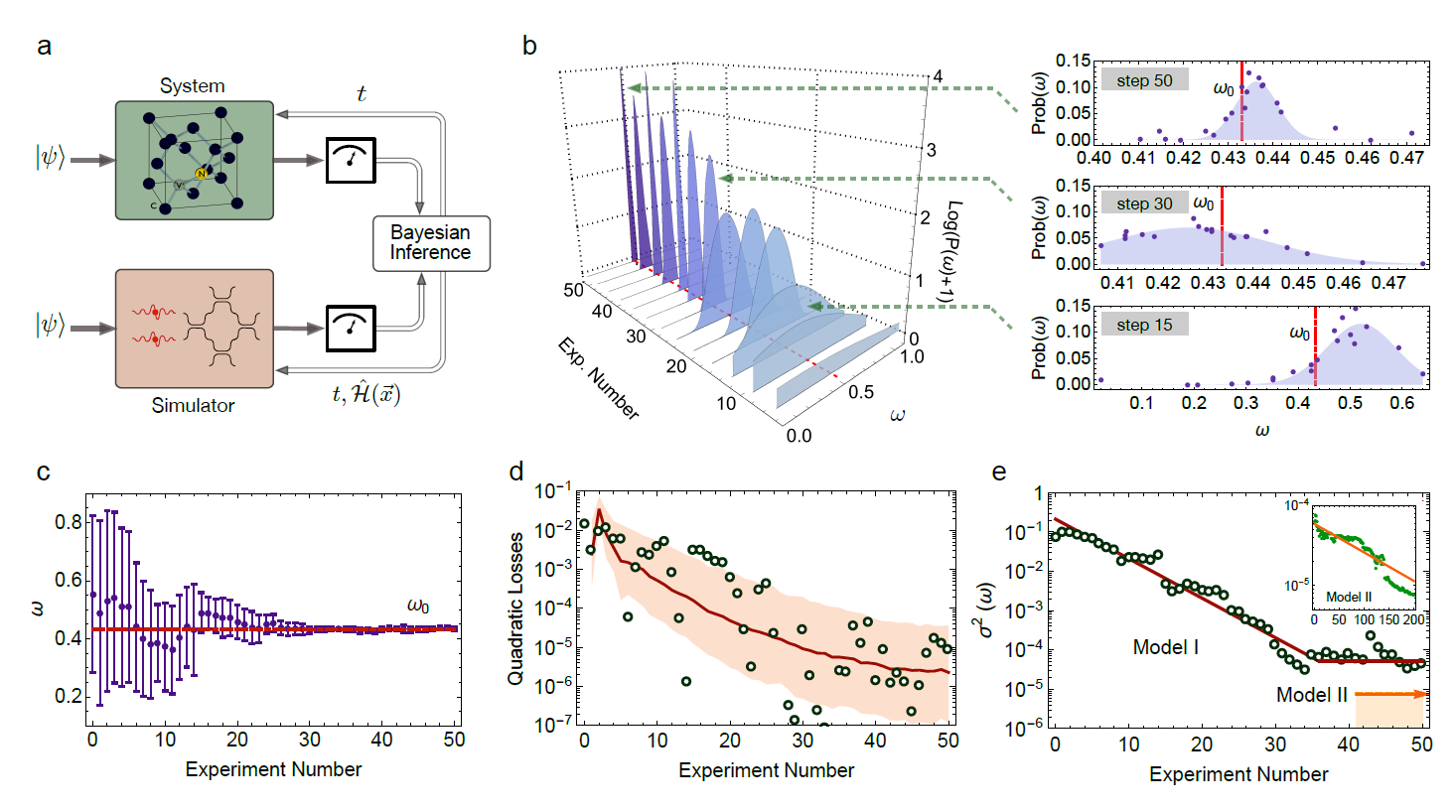}
\caption{\textbf{
Learning the electron spin dynamics on the quantum photonic simulator using QLE}. 
\textbf{a}, Schematic of QLE. 
The system Hamiltonian ${\hat H}(\vec{x}_0)$ (shaded green) is to be learned by a quantum simulator (shaded red) that embeds an abstract model $\hat{H}(\vec{x})$ of the target system. 
We here choose an electron spin in NV$^-$ centre as the target system and use a silicon photonic device as the simulator. 
In the system, the initial state $|\psi\rangle$ is evolved for a time $t$ and measured.  
The simulator mimics the system dynamics according to the model and obtains an estimate of the likelihood function using the outcomes from the system. 
The likelihoods are then used to infer the posterior distribution of the parameters $\vec{x}$ via Bayes' rule and to calculate the next step time $t$. 
\textbf{b}, The QLE progressive learning of the electron spin Hamiltonian parametrised by a rescaled Rabi frequency $\omega=f/\Delta f$. 
The distribution  over Hamiltonian parameter $\omega$ is described by a discrete approximation using 20 particles. 
Within 50 steps, the distribution converges to the correct  value $\omega_0$ (dashed red line). 
Insets: the distribution of particles after 15, 30 and 50 steps. Points are experimental data and shaded areas are un-normalized Gaussian fittings. 
\textbf{c}, Evolution of the mean and standard deviation of the distribution. 
Error bars are $ \pm 1$ s.d. of the distribution. 
\textbf{d}, Evolution of the quadratic losses.  
Circles are experimental data and the line represents theoretical simulation results with a $67.5 \%$ confidence interval (shaded area). The theoretical simulation was averaged over 500 runs of QLE. 
\textbf{e}, Model validation and improvement. 
The presence of other physical effects in the system that are not describable by the model $\hat{H}(\omega)$ (Model I) limits the amount of extractable information, as manifested by a saturation of the distribution variance at $\sigma^2(\omega)\simeq4.2\times10^{-5}$ after approximately 35 steps. 
The adoption of a new two-parameters model $\hat{H}'(\omega,\alpha)$ (Model II), which includes the presence of chirping, allows to achieve a covariance below $\left\lVert\Sigma\right\rVert_2=7.5\times 10^{-6}$ (the shaded area). 
Inset: covariance norm evolution of the Model II. 
}
\label{fig:DataQLE_NV}
\end{figure*}

Silicon quantum photonics is a promising platform for the realisation of manufacturable quantum technologies
~\cite{Joshreview,  Silverstone2015, Wang:16,Najafi:2015}.
Our silicon device integrates entangled photon generation, projective measurements, single-qubit and two-qubit operations onto a single chip, as shown in Fig.~\ref{fig:NVvsChip}b.
Photons are generated and entangled in the path-encoded state $(|0_s\rangle|0_i\rangle +|1_s\rangle |1_i\rangle)/{\sqrt{2}}$,
with $s$ and $i$ indicating signal and idler photons~\cite{Wang:16}. 
Then the idler photon is prepared in the state $|\psi_i\rangle$ and undergoes an arbitrary unitary evolution, $\hat U$ or $\hat V$, conditional upon the logical state of the signal photon~\cite{Zhou2011}. 
This entangled state
$(|0_s\rangle \hat{U}|\psi_i\rangle+|1_s\rangle \hat{V}|\psi_i\rangle)/{\sqrt{2}}$
is realised upon the coincidental detection of
the idler photon indicated by the blue dots, and the signal photon indicated by the red dots in Fig.~\ref{fig:NVvsChip}b.
The overlap between $\hat{U}|\psi_i\rangle $ and $\hat{V} |\psi_i\rangle$ is evaluated measuring the control qubit, enabling the estimation of the likelihoods for our QHL implementations (see Methods and  Suppl. Info.~1).

The solid-state spin-qubit dynamics~\cite{Jelezko2004, Togan:2010, Bernien, DirkNVPC, Chen2016} under test
are between the $m_\text{s}=0$ and $m_\text{s}=-1$ states of the electron ground-state triplet (Fig.~\ref{fig:NVvsChip}e) in the NV$^{-}$ centre.
Optical addressing, read-out, and microwave (MW) manipulation of the electron spin are performed with a bespoke confocal microscope arrangement.
At the transition frequency of \unit{2.742}{\giga\hertz},
the electron spin is optically initialised into the $m_\text{s}=0$ state.
The electron spin is then coherently driven in a single Rabi sequence (Fig.~\ref{fig:NVvsChip}f), for a given evolution time $t$, by applying MW pulses of a fixed but arbitrarily chosen power. 
The photo-luminescence (PL) indicating the spin state is detected and used to obtain the output probability (see Methods and Suppl. Info.~2).

The general aim of QHL is to find the parameters $\vec{x}_0$ that best describe the dynamical Hamiltonian evolution of the system via $\hat{H}_0=\hat{H}(\vec{x}_0)$.
Learning the Hamiltonian relies on an estimation of likelihoods, which can be exponentially hard to compute on classical machines. 
However, a quantum simulator can be programmed for a parametrised Hamiltonian $\hat{H}(\vec{x})$ such that the observed data allows the efficient estimation of its associated likelihoods. 
The first QHL protocol we implemented is called \textit{quantum likelihood estimation} (QLE). 
The initial state $|\psi\rangle$ of the target system is evolved for a time $t$ and measured in a basis $\{|D\rangle\}$, as shown in Fig.~\ref{fig:DataQLE_NV}a. The observed data $D$ is fed to the quantum simulator which simulates state evolution and measurement assuming $\hat{H}(\vec{x})$ as the true Hamiltonian. Given $\vec{x}$, the probability $\text{Pr}(D|\vec{x})=|\langle D |e^{-i\hat{H}(\vec{x})t}|\psi\rangle|^2$ of obtaining $D$ is known as the likelihood function for QLE. We then use $\text{Pr}(D|\vec{x})$ in combination with an approximate form of Bayesian inference known as Sequential Monte-Carlo algorithms (SMC) to learn $\vec{x}$ and estimate its uncertainty. In this approximation, a finite set of points in the parameters space $\{ \vec{x_i } \}$, called \textit{particles}, is used to describe the probability distribution (See Suppl. Info.~3. for details).

\begin{figure}[ht!]
\centering
\includegraphics[width=0.4\textwidth]{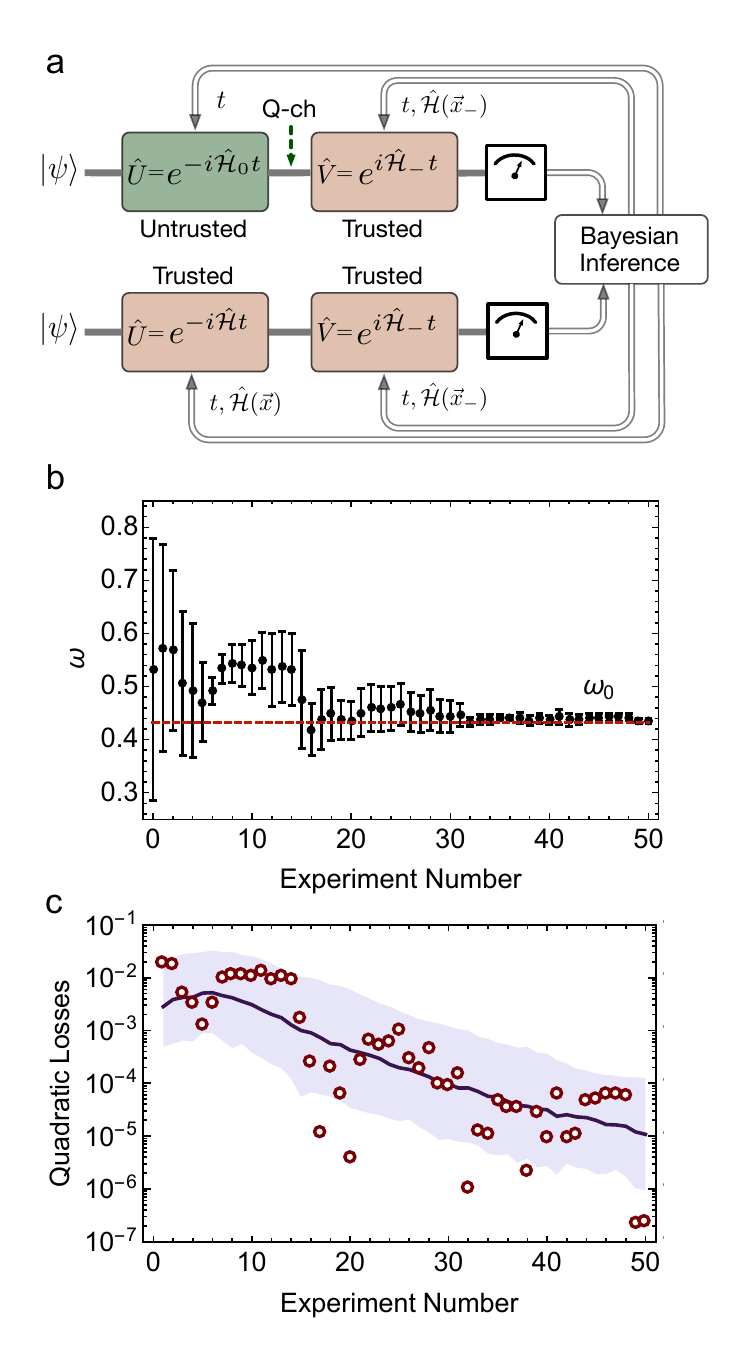}
 \caption{\textbf{Characterising the operation of the quantum photonic device using IQLE. 
} 
\textbf{a},
Schematic of IQLE. 
The untrusted quantum system is shaded green and the trusted quantum devices 
are shaded red. 
IQLE lies in the inversion of the evolution via a Hamiltonian $\hat{H}(\vec{x}_-)$ 
implemented with a trusted device (top one). The trusted and the untrusted are linked by a coherent quantum channel. 
The inversion is performed also in the likelihood estimation on the trust quantum devices (bottom one). 
Results are classically processed for Bayesian inference.
\textbf{b}, Evolution of the mean and standard deviation of the distribution of the rescaled frequency $\omega$, while IQLE is converging to the expected value of the $\omega_0$ (dashed red line). The determining of $\omega_0$ is equivalent to the characterising of $\hat{\sigma}_x$-rotation.  
Error bars are $ \pm 1$ s.d. of the distribution. 
\textbf{c}, Exponential decrease of quadratic loss for IQLE. Experimental data are shown as circles, and theoretical simulation data are shown as a line with a $67.5\%$ confidence interval (shaded area). 
The theoretical simulation was averaged over 500 runs of IQLE. 
 }
 \label{fig:IQLE}
 \end{figure}

Our silicon-photonics device and the NV$^{-}$ centre were interfaced with a classical computer, such that experimental data directly enabled QLE. 
Rabi oscillations of the NV$^-$ centre's electron spin can be modelled by a Hamiltonian of the form $\hat{H}(f)=\hat{\sigma}_x f/2$ acting on the initial state.
The silicon-photonics chip simulated the model $\hat{H}(f)$ to learn the Rabi frequency $f$ and to enable the calculation of the likelihood function for each particle.
At each step of the QLE implementation,
the evolution time $t$ was chosen adaptively for the NV$^-$ electron spin performing a single Rabi sequence. PL results were calculated from $3$ million iterations for each sequence.
The likelihoods obtained were then used to update the prior distribution via the classical computer, before proceeding to the next step.
The prior distribution $\text{Pr}(f)$ of the particles was initialised to be uniform between \unit{0} and an arbitrary value $\Delta f$, where we chose $\Delta f=100/2\pi ~ \mega\hertz$.
For clarity, we consider the rescaled quantity $\omega=f/\Delta f$ distributed in the interval $\omega\in[0,1]$. 

We performed QLE with 50 steps using a $20$ particles SMC approximation to learn the electron spin dynamics of the system. 
Figures~\ref{fig:DataQLE_NV}b,c show the particle's distribution converging to the correct value $\omega_0$. 
The final learned value corresponds to the Rabi frequency $f_{\text{ QLE}}=\unit{(6.93\pm 0.09)}{\mega\hertz}$, given by the mean and standard deviation of the distribution, which is consistent with the referenced value $f_0=
\unit{6.90}{\mega\hertz}
$ 
obtained with the fit of the Rabi oscillations measurements (see Suppl. Info.~2).
Thus the simulator successfully learns the parameter that best represents this Hamiltonian, without prior knowledge of the Rabi frequency. We note that the total number of measurements on the NV$^{-}$ system required for QLE is smaller than those for the fit ($\simeq 200$). 
The fast experimental convergence of the algorithm to $\omega_0$ is observed through the evolution of the quadratic losses of the particles distribution (here equal to the mean-squared error) achieving a final value of approximately $10^{-5}$ (see Fig.~\ref{fig:DataQLE_NV}d).

Figure~\ref{fig:DataQLE_NV}e shows that after an exponential decay in the first 35 steps, the variance $\sigma^2$ saturates at approximately $4.2 \times 10^{-5}$.
This saturation indicates that the algorithm stops learning within this model (Model I). Such saturations are easy to spot within a Bayesian framework, because $\sigma^2$ can be directly computed from the posterior distribution. This strikingly illustrates that QHL can estimate the limitations of the physical model used to describe the dynamics of the system.
Knowing when a model has failed affords us the opportunity to improve upon it.  
The present model was improved by introducing chirping, described by a time-dependent Hamiltonian  
 $\hat{H}'(f,\alpha;t)=\hat{\sigma}_x (f+\alpha t)/2$ (Model II), where $\alpha$ is a chirping constant. 
Including chirping allows the algorithm  to continue learning with an exponential decay of the covariance below $\left\lVert\Sigma\right\rVert_2 = 7.5\times 10^{-6}$ (see Fig.~\ref{fig:DataQLE_NV}e).
The final learned values of the two parameters are
$f_{\text{QLE}}=(7.00 \pm 0.04)~\text{MHz}$
and
$\alpha_{\text{QLE}}=(-0.26 \pm 0.04)~\text{MHz}^2$,
which are comparable with the values
$f_0=
6.94~\text{MHz}
$
and
$\alpha_0=-0.28
~\text{MHz}^2$
calculated with a full chirped Rabi fit (see Suppl. Info.~4).
A  formal comparison between the performances of the two models is given by the Bayes factor $K$, defined as the ratio of the average likelihoods calculated for each of the two models.
Considering all of the data collected from the NV$^{-}$ centre in performing the algorithm, we obtain $K=560$, which provides strong evidence in favour of the new model (despite its increased complexity).
This demonstrates that QHL not only estimates the best model parameters, but that it also instructs us to improve the model itself, providing potentially crucial insights into underpinning physical processes (see Suppl. Info.~4).

Though QLE is scalable, it often requires short evolution times to ensure the likelihood evaluation is tractable, which can preclude exponential reductions in the number of experiments needed, and makes the SMC approximation more error prone.
Yet if it is possible to couple two quantum devices via a quantum (rather than a classical) channel, such as photon-NV spin coupling systems~\cite{Togan:2010} or different gates on a single photonic chip~\cite{Carolan:15}, 
an \textit{interactive quantum likelihood estimation} (IQLE) algorithm can be adopted to overcome these problems~\cite{Wiebe2014,Wiebe2014b}. 

Similar to QLE, in IQLE the state initially evolves forward in time with the Hamiltonian of the system $\hat{H}(\vec{x}_0)$. However, the transformation is then inverted by the time-reversed Hamiltonian evolution $\hat{H}_-=\hat{H}(\vec{x}_-)$, with $\vec{x}_-$ sampled from the prior distribution (Fig.~\ref{fig:IQLE}a). To ensure the backwards transformation via $H_-$, the state must be transferred from the system to the simulator. Thus IQLE requires the presence of a coherent quantum channel between them.
IQLE enables a number of significant features. It has been shown that the likelihood function for the two-outcome experiments, which involves computing $|\langle \psi | e^{i\hat{H}_- t} e^{-i\hat{H}(\vec{x})t}|\psi \rangle|^2$, is efficient for $\hat{H} \approx \hat{H}_-$ even if $\|\hat{H}\|t \gg 1$~\cite{Wiebe2014}. IQLE is also expected to be much more resilient to errors in the inference process, making it more robust for experimental implementations and critical device verifications~\cite{Wiebe2014b}.

The required quantum channel naturally exists in a single quantum device. In this case, IQLE can be applied to use calibrated gates to efficiently characterise other un-calibrated gates on the same quantum device, which now respectively represent the trusted hardware and the untrusted system to be validated. This application illustrates how IQLE can be used to help characterise and verify quantum devices, improving the scalability in many-qubit systems in which characterisation will be a key challenge. 

We implement IQLE entirely on the photonic chip, showing its ability to characterise single-qubit operations of quantum devices. 
In our experiment the photonic device plays the role of both the untrusted system and the trusted hardware, allowing the natural implementation of a quantum device self-verification. 
The operation to be characterised here is $e^{-i f_0 t  \hat{\sigma_x} /2}$, where $f_0$ matches the value of the fitted Rabi frequency, chosen for consistency with the previous QLE demonstration.
Thus characterising this $\hat{\sigma}_x$-rotation operation is equivalent to learning the Rabi frequency. 
Similar to the QLE demonstration, the Hamiltonian $\hat{H}(f)$ of Model I was simulated to learn the parameter $\omega=f/\Delta f$. 
In each step of IQLE, the experiment was implemented twice: once for measuring the outcomes from the untrusted $\hat{\sigma}_x$-rotation (top part in Fig.~\ref{fig:IQLE}a ), and once for estimating the likelihoods (bottom part in Fig.~\ref{fig:IQLE}a). 
See Methods for more details. 
Figure \ref{fig:IQLE}b shows the experimental results for the estimated $\omega$ as given by the posterior mean and standard deviation at each step of IQLE. 
The particle distribution converges quickly to the correct value $\omega_0$.  
After 50 algorithm steps we obtain $f_{\text{ IQLE}}=\unit{(6.92\pm 0.08)}{\mega\hertz}$, which is within $1$~s.d.
of the implemented Rabi frequency $f_0=
\unit{6.90}{\mega\hertz}
$.  
The evolution of the quadratic losses  (Fig. \ref{fig:IQLE}c) indicates that the parameter is learned exponentially fast, with a final quadratic loss value of approximately $10^{-7}$.  The convergence of the algorithm to the implemented value $\omega_0$ indicates the successful self-verification of the quantum device.

We report the first demonstration of QHL showing the capability of validating Hamiltonian models and verifying quantum devices.
While these experiments use a digital quantum photonic simulator for the  demonstration, QHL is universal and can be implemented on any quantum computing platform (e.g. ~\cite{Carolan:15, Debnath2016, Barends}). 
Furthermore, this learning protocol applies to non-digital simulators, which is particularly of interest when certain classes of analogue quantum simulations are likely to approach a regime beyond that available to classical supercomputers in the medium term~\cite{Parsons1253, Labuhn:2016}.  
With anticipated future developments in quantum hardware, the QHL protocol can be scaled up to learn more complex Hamiltonians, and promises the early delivery of quantum-enhanced computational techniques to efficiently characterise and verify quantum systems and technologies, and to investigate foundational physics. 

\section*{Methods}
{\small
\noindent\textbf{Diamond NV$^{-}$ centre and setup.} 
The bulk diamond hosting the negative changed NV$^-$ centre is a chemical vapour deposition (CVD) grown sample (electronic grade) with a natural abundance of $1$ ppb nitrogen impurities, see Inset in Fig\ref{fig:NVvsChip}c. The NV$^{-}$ centre was positioned in the static magnetic field at room temperature. 
With the help of optical detected magnetic resonance (ODMR), we perfectly aligned a small external magnetic field of $5\milli\tesla$ 
in the direction of the NV$^{-}$ centre's axis, lifting the degenerated $m_\text{s}=\pm 1$ spin states.
 Fig. S3a shows the ODMR of the NV$^-$ centre used in the experiment, which was scanned under continuous optical laser and microwave (MW) excitation, indicating the  transition from $m_\text{s}=0$ to $m_\text{s}=-1$. \\
\noindent \textbf{Silicon quantum photonic device and setup.} 
The quantum device was manufactured on the standard Silicon-on-Insulator (SOI) wafer using \unit{248}{\nano\meter}-ultraviolet photolithography and reactive-ion etching. Single photons were generated and guided in silicon waveguides with a cross-section of \unit{450} {\nano\meter} $\times$ \unit{220} {\nano\meter}. 
The single photon-pair sources were designed with a \unit{1.2} {\centi\meter} length. 
The relative phase between different paths was manipulated using thermo-optical phase-shifters obtained by metal deposition of titanium upon the silicon waveguides. MMIs couplers were used as beam-splitters with near a 0.5 reflectivity. 
The device was wired-bounded on a PCB and each phase-shifter was individually controlled by an electronic driver with 12 bits resolution. 
A classical computer was used to process the photon statistics obtained through a time interval analyser from the quantum device, and perform the Bayesian inference to update the Hamiltonian model. \\
\noindent \textbf{Estimation of the likelihoods on the photonic chip.} 
The likelihood estimation for QLE requires the inner product between the evolved state $e^{-i \hat{H}(\vec{x}) t}|\psi\rangle$ and the state $|D\rangle$. In this work we have used the initial state of the NV$^-$ centre: $|D\rangle=|\psi\rangle=|m_s=0\rangle$. 
In order to calculate the inner product in our photonic device we exploit an entanglement-based technique~\cite{Cai2015}. The scheme realised in the integrated device allow us to produce  entangled states $(|0_s\rangle \hat{U}|\psi_i\rangle+|1_s\rangle \hat{V}|\psi_i\rangle)/{\sqrt{2}}$, as described in the main text. When performing a projective measurement on the $\hat{\sigma}_x$ eigenbasis 
$\{|+\rangle,|-\rangle\}$ of the signal qubit,  
we obtain
\begin{equation}\label{RePart}
\text{Re}(\langle\psi|\hat{U}^\dagger \hat{V}|\psi\rangle)=2 p_+ - 1
\end{equation}
where $p_{+}$ is the probability to get the outcome $|+\rangle$. Similarly, when performing a projective measurement on the $\hat{\sigma}_y$ eigenbasis $\{|+i\rangle,|-i\rangle\}$, 
we obtain 
\begin{equation}\label{ImPart}
\text{Im}(\langle\psi|\hat{U}^\dagger \hat{V}|\psi\rangle)=2 p_{+i} - 1
\end{equation}
where $p_{+i}$ is the probability to get the outcome $|+i\rangle$. The likelihood for QLE is given by $\mathcal{L}_{\text{QLE}}=|\langle \psi|e^{-i\hat{H}(\vec{x})t}|\psi \rangle|^2$, which can be easily obtained setting $\hat{U}=\hat{\mathds{1}}$ and $\hat{V}=e^{-i\hat{H}(\vec{x})t}$. Using  Eq.~\ref{RePart} and Eq.~\ref{ImPart}, we have
\begin{equation}\label{likelihood}
\mathcal{L}_{\text{QLE}}=(2 p_{+} - 1)^2+(2 p_{+i} - 1)^2
\end{equation}
The values of $p_{+}$ and $p_{+i}$ are calculated performing the single-qubit operations on the control qubit and measuring photon coincidences. \\
For IQLE the likelihood $\mathcal{L}_{\text{IQLE}}=|\langle \psi|e^{i\hat{H}(\vec{x}_{-})t}e^{-i\hat{H}(\vec{x})t}|\psi \rangle|^2$ is obtained similarly using Eq.~\ref{likelihood} but setting $\hat{U}=e^{-i\hat{H}(\vec{x}_{-})t}$ and $\hat{V}=e^{-i\hat{H}(\vec{x})t}$. 
The quantum channel required for IQLE in our case is thus given by the entanglement generated in the sources. We remark that using this scheme to implement IQLE all evolutions are forward in time, in contrast with the original approach where the time reversal $e^{i\hat{H}(\vec{x}_{-})t}$ is performed by a backwards evolution in time~\cite{Wiebe2014}. This can make our entanglement-based approach amenable for analogue quantum simulators. However, it comes at the cost of additional entanglement between the system and an ancillary qubit.
}

\printbibliography

\section*{Acknowledgements}
We thank J. W. Silverstone, D. Bonneau, X. Zhou, D. P. Tew, J. Carolan, S. Jia and C. Granade for useful discussions.  
We thank K. Ohira, N. Suzuki, H. Yoshida, N. Iizuka and M. Ezaki for the device fabrication. 
We acknowledge the support from the
Engineering and Physical Sciences Research Council (EPSRC), European Research Council (ERC), including Photonic Integrated Compound Quantum Encoding (PICQUE), 
Beyond the Barriers of Optical Integration (BBOI), 
Quantum Simulation on a Photonic Chip (QuChip), 
and Wavelength-tunable Advanced Single Photon Sources (WASPS, grant agreement no.\  618078), US Army Research Office (ARO), and the Centre for Nanoscience and Quantum Information (NSQI). 
M.P. acknowledges the support by a STSM Grant from COST action MP1403.  
J.L.O'B. acknowledges a Royal Society Wolfson Merit Award and a Royal Academy of Engineering Chair in Emerging Technologies.
J.G.R. is sponsored under EPSRC grant no.\ EP/M024458/1.
A.L. acknowledges support from an EPSRC Early Career Fellowship.
M.G.T. acknowledges the support from an EPSRC Early Career Fellowship and the Toshiba Research Fellowship scheme. 

\section*{Author contributions}
J.W., S.P. and R.S. contributed equally to this work.
J.W., S.P., R.S., S.K. and N.W. conceived and designed the experiments. N.W. provided theoretical support. S.P. and M.P. performed the simulations. 
J.W., S.P., R.S., S.K. and A.A.G built the setups, carried out the experiment, analysed the data and wrote the manuscript with feedback from all authors. N.W., J.L.O'B., J.G.R., A.L. and M.G.T. managed the project. 

\end{document}